\documentstyle[12pt]{article}
\include{epsf}
\def\lo{\langle 0 |}

\def\gmmu{\gamma _{\mu}}

\def\atop{ \frac{ \alpha_{s}}{4 \pi} G_{\mu \nu}
 \tilde{G}_{\mu \nu} }
\def\atopa{\frac{ \alpha_{s}}{4 \pi} G_{\mu \nu}^{a}
 \tilde{G}_{\mu \nu}^{a} }

\def\fc{ f_{\eta'}^{(c)} }
\def\gmf{\gamma _{5}}

\def\la{\langle }
\def\ra{ \rangle }

\def\er{ | \eta' \rangle}

\newcommand{\beq}{\begin{equation}}
\newcommand{\eeq}{\end{equation}}
\newcommand{\bea}{\begin{eqnarray}}
\newcommand{\eea}{\end{eqnarray}}

\begin{document}
                                        \begin{titlepage}
\begin{flushright}
hep-ph/9706251
\end{flushright}
\vskip1.8cm
\begin{center}
{\LARGE
 Polarized Intrinsic Charm  \\      
\vskip0.7cm
as a Possible Solution to the Proton Spin \\
\vskip1.0cm
Problem
}         
\vskip1.5cm
 {\Large Igor~Halperin} 
and 
{\Large Ariel~Zhitnitsky}
\vskip0.5cm
        Physics and Astronomy Department \\
        University of British Columbia \\
        6224 Agriculture Road, Vancouver, BC V6T 1Z1, Canada \\  
        {\it and}\\
        Isaac Newton Institute For Mathematical Sciences\\
        20 Clarkson Road, Cambridge, CB3 0EH, U.K.\\
     {\small e-mail: 
higor@physics.ubc.ca \\
arz@physics.ubc.ca } \\
PACS numbers: 12.39.-t, 12.38.Aw, 11.15.Pg, 12.38.Lg .  \\
Keywords: Low energy theorems, U(1) problem, B decays, instantons.
\vskip1.5cm
{\Large Abstract:\\}
\end{center}
\parbox[t]{\textwidth}{ 
We argue that a large fraction of the proton spin comes from 
a contribution due to a nonperturbative intrinsic charm component 
of the proton. An account of this contribution removes an apparent 
contradiction between the data and exact K\"{u}hn-Zakharov low 
energy theorem. On the other hand, we show that a large intrinsic 
charm spin component of the proton is implied by recent CLEO data on 
$ B \rightarrow \eta' $ decays. We argue that the proton spin and 
$ B \rightarrow \eta' $ data are manifestations of the same physics 
with full agreement between two different estimates of the intrinsic 
charm component of the proton spin. Our results suggest an explanation 
of the polarized DIS data and have profound consequences for future 
experiments on charm production off a polarized proton. A  microscopic  
origin of the effect is related to strong well-localized gluon 
fluctuations (instantons) where the mass of the charmed quark is not a suppression factor at all, as was  naively expected earlier.}

\vspace{1.0cm}

                                                \end{titlepage}

\section{Introduction}

The famous proton spin problem is now almost ten years old. 
A wide interest in the structure of the proton
spin was triggered by the 1987
EMC experiment \cite{EMC} which has measured the first 
moment of the proton spin structure function $ 
\Gamma_{1}^{p} \equiv \int dx g_{1}^{p}(x) = 0.126 \pm 0.018
$. This result implies that the polarized deep inelastic 
scattering (DIS) data cannot 
be explained simply by the valence quark spin components
\cite{EJ}, but, in addition,
 a substantial sea-quark polarization is needed.
 Such a polarized quark 
sea seems very miraculous from the point of view of both the 
relativistic quark model (where valence quarks account
for 3/4 of the proton spin) and the parton model (where the 
helicity conservation prohibits a perturbative sea polarization 
by hard gluons). This problem, sometimes called ``the proton spin 
crisis", has stimulated a great deal of interest over the last decade
(see e.g. \cite{Ans} for a review and 
representative list of references).

Nowadays, it is commonly accepted that these are 
gluons that play an important role in the spin structure of the 
nucleon.    
Depending on a factorization scheme, a gluon contribution to  
$ \Gamma_{1}^p $  is either given by an independent anomaly term 
in the chiral-invariant 
scheme \cite{anomspin}, or included in spin-dependent quark
distributions in the Operator Product Expansion (OPE) 
based gauge-invariant scheme advocated 
by Jaffe and Manohar \cite{JafMan}. 
In the Jaffe and Manohar approach, which we will
follow, the data are usually interpreted 
as the manifestation of a negatively polarized
$s$-quark component in the proton. An equivalence of these two 
descriptions for the first 
moment $ \Gamma_{1}^p $  was established by Bodwin 
and Qiu \cite{BQ}. 
While the former presumably makes more contact
with one's intuition and the parton model, the latter is 
based entirely on QCD and turns out more convenient technically
(see the Cheng's paper in \cite{Ans} and below).

Irrespective of the question of interpretation of  
$ \Gamma_{1}^p $, the gluon polarization $ \Delta G(x) $ 
in the proton is a very interesting object
on its own. This quantity is currently attracting a great deal
of attention of both theorists and experimentalists. Among 
experimental tests on  $ \Delta G(x) $  a hadronic heavy-quark
production (and, in particular, the charm production) is believed
to be the best one. The reason for this belief is that within 
the standard perturbative photon-gluon fusion prescription
the charm production amplitude is very sensitive to the gluon 
distribution in the nucleon. Among future experiments, 
intended to probe the 
gluon polarization through this scenario, one should mention
the RHIC and HERA-$\vec{N}$ projects on a charm production
in polarized $ p \bar{p} $ collisions, and the COMPASS experiment   
at LHC on an open charm production in polarized DIS (see the 
review by Ramsey in \cite{Ans} for references and more detail).
Existing theoretical expectations for these experiments are all
based on the perturbative photon-gluon fusion scenario.

However, while there is little doubt that this mechanism is 
operative for unpolarized DIS, its relevance may (and must)
be questioned in the case of polarized DIS. Indeed, in the latter 
case one deals with the axial channel instead of the vector one,
as it was for usual DIS. On the other hand, it is very well
known that physics of axial channels is drastically different
from that of vector ones. Unlike the latter,
{\bf axial and pseudoscalar channels are strongly influenced by 
nonperturbative effects.} The best known example of this difference 
is provided by a nonperturbative breakdown of the Zweig rule
in the pseudoscalar nonet. Thus, the perturbative charm production
mechanism, valid for usual DIS, certainly cannot be taken for
granted in the polarized case. In particular, strong 
nonperturbative fluctuations in the axial channel could induce 
a sizeable ``intrinsic charm" fraction of the proton spin,
thus substantially changing theoretical predictions for charm
production in polarized experiments.

The purpose of this paper is to argue that the latter hypothetical
situation is in fact precisely what happens in reality. 
We will
present two independent estimates of the 
intrinsic charm component
of the proton spin, which agree with each other and both indicate
that this quantity is quite large.
Our main argument is based on recent experimental results, while 
a second estimate relies on a theoretical result which has been known
for a long time, but unfortunately has not received much attention
in the literature. 
As will be shown
below, the observation of a large $c$-quark 
contribution to the proton spin turns out to be crucial 
for understanding the polarized DIS data. 
Our arguments go 
beyond perturbation theory. 

Our main estimate comes from  recent data of B-physics
(!), which at first sight might seem entirely uncorrelated 
with the proton spin problem. We mean the new CLEO results 
on $ B \rightarrow \eta' $ decays \cite{CLEO}. Recently, we have 
shown \cite{1} that $ \eta' $ production in B decays can only be 
explained by the presence of a large intrinsic charm component
$ \lo \bar{c} \gmmu \gmf c \er $ of the $ \eta' $. Moreover, a 
theoretical prediction for this quantity \cite{1} yields a good 
agreement with the data in both the exclusive \cite{1} and 
inclusive \cite{2} cases. On the other hand, it is fairly obvious
that this property of the $ \eta' $ implies a sizeable 
nonperturbatively generated intrinsic charm component of the proton
spin $ \la p | \bar{c} \gmmu \gmf c | p \ra $, as soon as there
exists a Goldberger-Treiman type relation between these matrix 
elements. 
A relation with the $B$-physics provides us 
with an estimate of this quantity which, as will be shown 
below, agrees with an estimate based on a pure theoretical 
result which we are about to discuss.

Our second estimate is based on a comparison of the DIS data with a
low energy theorem due to K\"{u}hn and Zakharov (KZ) \cite{KZ},
which is an {\it exact} (in the chiral limit) formula for the 
proton matrix element of the topological density. This important
result, obtained in 1990, sank into oblivion in subsequent 
publications. We understand that one of the reasons for this 
attitude was an apparent sign contradiction 
between the KZ formula
and the data. As we will argue, this contradiction is actually
removed by the intrinsic charm component of the proton 
spin in a very natural way and in full agreement with our
previous estimate based on the analysis of $B$ decays.
Therefore, the two independent lines of reasoning 
lead to the same conclusion on a role of the 
polarized charm in the 
proton spin.

We thus arrive at a coherent and attractive explanation of the 
polarized DIS phenomenology. In our opinion, the measured first 
moment $ \Gamma_{1}^p $ is determined (we assume the 
chiral limit) by four cornerstones:\\
(1) spontaneous breaking of the $ SU_{L}(3) \times SU_{R}(3)
$ chiral invariance (see below), \\
(2) a resolution 
of the $ U(1) $
problem, \\
(3) the proton matrix element of the topological
density, fixed by the KZ theorem, and \\
(4) the intrinsic charm component
of the proton spin, which is extracted from  data on 
 $ B \rightarrow \eta' $ decays.

 As we will show below, a consistent 
treatment of these 
four ingredients leaves very little (if any) of the 
mystery of $ \Gamma_{1}^p
$, and has profound consequences for the future experiments. Our 
presentation is organized as follows. We start in Sect.2 with 
a re-interpretation of experimental facts on the polarized DIS.
The only difference from the standard analysis is that we assume
a non-zero (but unspecified any further at this stage) value of the
matrix element $ \la p | \bar{c} \gmmu \gmf c | p \ra $ which 
describes the intrinsic charm component of the proton spin.
We 
use the data to select a $ SU(3) $ flavor singlet contribution
(which is a combination of the anomaly and intrinsic charm terms)
to different flavor components of the proton spin.
 We thus interpret
the data as a constraint on this $ SU(3) $ singlet contribution.
In Sect.3 we present the KZ theorem and then use it to select from
this combination a contribution of the intrinsic charm proper.  
In Sect.4 the same quantity is estimated independently from the 
data on $ B \rightarrow \eta' $ decays. In Sect.5 we 
briefly discuss consequences
 of our results for a charm production in polarized experiments. 
Our conclusions are presented in final Sect.6. 
    
\section{ SU(3) singlet proton spin from the data}
 
We start with the OPE analysis
of the first moment of the polarized $ g_1 $ structure function.
The only deviation from the standard treatment discussed in 
detail in 
\cite{Ans} is that we keep the charm component of the 
electromagnetic current of virtual photon. Retaining this 
term amounts to keeping trace of an intrinsic
charm component of the proton spin. In the standard analysis
this component is discarded on the grounds that
this contribution is expected to appear only at higher 
orders in $ \alpha_s $ from loop effects. As will be shown
below, this term is actually $ O(\alpha_{s}^0) $,
though it is suppressed by $1/m_c^2$ in agreement
with general arguments by Jaffe and Manohar \cite{JafMan}. 
Therefore,
we insist on keeping it in the analysis. On the 
other hand, at EMC energies
$ Q^2 \simeq 10 \; GeV^2 $, the mass of a struck $c$-quark 
can be neglected,
and thus the sole effect of taking the charm contribution into
account in the OPE analysis\footnote{Usual unpolarized DIS on 
intrinsic charm was considered in \cite{Ing}.}
is simply reduced to adding a 
corresponding charm contribution to light $ u,d,s $ quark
spin components. In the standard nomenclature, we calculate
the quantity  
\beq
\label{def}
\Gamma_{1}^{p} (Q^2) \equiv \int_{0}^{1} dx g_{1}(x, Q^2) \; .
\eeq
Using the OPE for a T-product of two electromagnetic 
currents and selecting an antisymmetric in Lorentz indices  
contribution, one arrives at the result
\beq
\label{OPE}
\Gamma_{1}^{p} (Q^2) = \frac{1}{2} \left( \frac{4}{9}
\Delta u (Q^2) + \frac{1}{9} \Delta d (Q^2) + 
\frac{1}{9} \Delta s (Q^2) + \frac{4}{9} \Delta c (Q^2)
\right) ( 1 - \frac{\alpha_{s}(Q^2)}{ \pi} + \ldots) \; .
\eeq    
Here $ \Delta q $ ( where $ q $ is any of the quark flavors)
stands for the spin component of the proton due to 
the quark flavor $ q $
\beq
\label{3}
s_{\mu} \Delta q(Q^2) = \la p| \bar{q} \gmmu \gmf q | p \ra \; ,
\eeq
($ s_{\mu} $ is the proton spin vector) and the 
operator in (\ref{3}) is normalized at the normalization 
point $ Q^2 $. To select a $ SU(3) $ singlet contribution
to (\ref{OPE}), it is convenient to use the combinations 
\beq
\label{4}
g_{A}^3 = \Delta u - \Delta d  \; , \; 
g_{A}^8 = \Delta u + \Delta d - 2 
\Delta s \; , 
\eeq
which have no anomalous dimensions and thus can be found 
from low energy neutron and hyperon beta decays data. Assuming the
$ SU(3) $ flavor symmetry, the non-singlet couplings are 
\beq
\label{5}
g_{A}^3 = F + D \;  , \;  g_{A}^8 = 3 F - D  \; . 
\eeq
For the values $ F = 0.459 \pm 0.008 \; , \; D = 0.798 \pm 0.008
$, this yields $ g_{A}^8 = 0.579 \pm 0.025 $ \cite{Ans}. 
Using relations (\ref{4}), one can now express the 
answer for $ \Gamma_{1}^p $ in terms of the (known) constants
$ g_{A}^3 , g_{A}^8 $ and the unknown SU(3) singlet combination
\beq
\label{6}
\Delta \Sigma(Q^2) \equiv \Delta u (Q^2)+ \Delta d (Q^2)
+ \Delta s (Q^2) 
\eeq
plus an intrinsic charm part $ \la p | \bar{c} \gmmu \gmf c | p \ra
$. We obtain
\beq
\label{7}
\Gamma_{1}^{p} = C_{NS}(Q^2) \left( \frac{1}{12} g_{A}^3 + 
\frac{1}{36} g_{A}^8 \right) + \frac{1}{9} C_{S} (Q^2) \left( 
\Sigma (Q^2)+ 2 \Delta c (Q^2) \right) \; .
\eeq
As we have said before, the only difference from the standard formulas
in Eq.(\ref{7}) is the presence of an intrinsic charm term $ 
\Delta c $. Coefficients $ C_{NS}, C_{S} $ account for perturbative 
QCD corrections $ C_{NS (S)} = 1 - \alpha_s / \pi + \ldots $ which
can be found in \cite{Ans}. 

A global fit to all available data (including the data of SMC, 
E142 and E143) together with a treatment of higher order QCD 
corrections, but neglecting the charm contribution 
gives\footnote{These 
numbers were slightly 
updated by Ellis and Karliner in \cite{Ans}. To our accuracy
this difference is unessential.}
\cite{EK}
\beq
\label{8}
\Delta u = 0.83 \pm 0.03 \; , \; \Delta d = - 0.43 \pm 
0.03 \; , \; \Delta s = - 0.10 \pm 0.03  
\eeq
with 
\beq
\label{9}
\Delta \Sigma = 0.31 \pm 0.07
\eeq
at $ Q^2 = 10 \; GeV^2 $. In our case a difference 
from this result comes due to retaining the intrinsic 
charm $ \Delta c $ term in Eq.(\ref{7}). Therefore, we translate 
Eq.(\ref{9}) in the constraint
\beq
\label{10}
g_{A}^0 \equiv \Delta \Sigma + 2 \Delta c = 0.31 \pm 0.07 \; .
\eeq
(again $ Q^2 = 10 \; GeV^2 $ is implied). It is convenient at this 
stage to proceed to derivatives instead of axial currents which stand
in (\ref{10}). Using the result of the OPE for the $c$-quark bilinear
(see Eq.(\ref{new1}) below) and working in the chiral $ SU(3) $
limit (which will be implied in what follows), we obtain for
$ n_f = 3 $ light flavors
\beq
\label{11}
\la p | n_{f} \frac{\alpha_s}{4 \pi} G \tilde{G} - 
2 \cdot \frac{1}{16 \pi^2 m_{c}^2 } g^3 G \tilde{G} G | p 
\ra =  g_{A}^{0} \cdot 2 M_{p} \, \bar{p} i \gmf p \; . 
\eeq
Here the first term is the standard contribution
due to the light quarks. The second term is new, and plays
a very important role in what follows. It originates in 
the $c$-quark contribution $ \Delta c $, and technically 
comes from the derivative of the axial charmed current
\bea
\label{new1}
\partial_{\mu}(\bar{c}  \gmmu\gmf c) &=& \atopa +  
2 m_c \bar{c} i \gmf c  \\ 
&=& \atopa  - \atopa - \frac{1}{16 \pi^2
 m_{c}^2 }  
g^3 f^{abc} G_{\mu \nu}^a \tilde{G}_{\nu \alpha}^b 
G_{\alpha \mu}^c  + \ldots  \nonumber
\eea
(see the appendix in \cite{1} for a detailed derivation 
of this result).

If we were to forget about the 
second charm-induced term in Eq.(\ref{11}), we would 
conclude that
the quark contribution to the proton spin is given by 
the proton 
matrix element of the topological density (which
has to be understood as the limit of a near-forward matrix 
element for vanishing momentum transfer)\footnote{
While, as will be 
explained
below, this neglect is not legitimate, our point here is different. 
The fact that the light quark fraction of the proton spin
$ \Delta \Sigma $ is {\it equivalent} to a particular 
value of the anomaly matrix element over the nucleon was strongly
emphasized by Jaffe and Manohar \cite{JafMan}. This statement is 
a consequence of the exact equation of motion. A very similar
situation holds in another anomalous case: the $ \eta' $ residue
$ f_{\eta'} $ measures a coupling of the $ \eta' $ to quarks, 
$ \lo \sum \bar{q}_{i} \gmmu \gmf q_{i} \er = 
i\sqrt{3} f_{\eta'} q_{\mu}$.
On the other hand, the same parameter gives in the chiral
limit a coupling of the 
 $ \eta' $ to gluons: $ \lo n_f\alpha_s/(4 \pi) G \tilde{G} \er = 
\sqrt{3}f_{\eta'} m_{\eta'}^2 $ as a consequence of exact equation of 
motion.  As was stressed in \cite{JafMan}, gluon operators cannot
appear explicitly in the OPE (\ref{OPE}) since there is no gauge
invariant gluon operator of spin 1 and twist 2. Thus, 
in the gauge invariant
OPE approach to polarized DIS a gluon contribution is ``hidden"
(via the subtraction of regulator terms) 
in quark spin dependent distributions, which therefore do not have 
a parton interpretation. Later, Bodwin and Qiu have demonstrated
\cite{BQ} that any gauge invariant (i.e. Pauli-Villars) 
regularization of 
the photon-gluon scattering diagram automatically yields zero 
contribution of this diagram to the first moment $ \Gamma_{1}^{p}
$, in agreement with the general logic of OPE. Moreover, the 
approach based on the OPE is more convenient technically than
the alternative method \cite{anomspin}, since it deals at every step
with matrix elements of local gauge invariant operators, which are 
well defined objects. }. 
The corresponding matrix element is fixed by 
a beautiful result due to K\"{u}hn and Zakharov (KZ) \cite{KZ}, 
which provides us with an {\it exact} (in the chiral limit)
answer for the matrix element of the topological density:
 \beq
\label{new2}
\la p | n_{f} \frac{\alpha_s}{4 \pi} G \tilde{G} | p \ra 
= - \frac{2n_f }{ 3 b } 2M_{p} \, \bar{p} i \gmf p .
\eeq
As will be shown below, this elegant formula is not sufficient 
to explain the polarized DIS data. The second charm-induced term 
in Eq.(\ref{11}) turns out to be absolutely crucial in  this problem.
We postpone a corresponding discussion
to the next section, while here 
we want to go one step further and ask the question whether it is 
possible to find  flavor contributions 
to the proton spin separately. The answer 
to this question is affirmative. To make contact with previous 
analyses which have not taken into account the charm contribution, 
we find it convenient to split it equally between the three light 
flavor contributions, i.e. to consider ``shifted" values
$ \Delta q' \equiv \Delta q + 2/3 \Delta c $ for $ q = u,d,s $. 
This is in accordance with our definition (\ref{10}) where
$\Delta c $ simply redefines the singlet contribution 
$\Delta\Sigma$.
Such a procedure 
amounts to adding an 
$ SU(3) $
singlet piece equally to each of the spin dependent light quark 
distributions, which 
was accounted previously by   the 
  anomaly term alone without a charmed contribution.
   We may then translate the result of  
Ellis-Karliner fit (\ref{8}) in the values of $ \Delta q' $ rather
than $ \Delta q $ as it was 
 originally stated in Eq.(\ref{8}). On the other hand, taking 
the derivative for each of the three flavor $ q = u,d,s $, we obtain
\beq
\label{12}
\la p| 2 m_q \bar{q} i \gmf q + \tilde{Q} | p \ra
=  \Delta q' \cdot 2 M_{p}  \, \bar{p} i \gmf p \; ,
\eeq
where $ \tilde{Q} $ stands for the $ SU(3) $ singlet part : 
\beq
\label{13}
\tilde{Q} = \frac{\alpha_s}{4 \pi} G \tilde{G} - \frac{2}{3} 
\frac{1}{16 \pi^2 m_{c}^{2} } g^3 G \tilde{G} G \; .
\eeq
Here we would like to remind the reader that we work in the chiral 
limit $ m_{q} = 
0 $. However, it would be completely erroneous to 
put $ m_{q} = 0 $ 
in matrix elements (\ref{12})
 from the very beginning. Proceeding this way, we would 
miss the phenomenon of spontaneous breaking of the 
chiral $ SU(3)_{L}
\times SU(3)_{R} $ invariance. Goldstone bosons, resulting
from this breakdown, have masses $ m^2 \sim m_q $, and therefore
are the only surviving intermediate states in matrix elements $ 
\la p | m_q \bar{q} i \gmf q | p \ra $ in the chiral limit $ m_q 
\rightarrow 0 $, which is only taken after
a saturation of these matrix elements by goldstones.
Technically, it is more convenient not to do it explicitly, 
but rather use relations (\ref{4}). Taking the derivatives, we 
obtain 
\bea
\label{14}
\la p | 2 m_u \bar{u} i \gmf u - 2 m_d \bar{d} 
i \gmf d | p \ra
&=& g_{A}^{3} \cdot 2 M_{p} \, \bar{p} i \gmf p \; , \nonumber \\
\la p | 2 m_u \bar{u} i \gmf u + 2 m_d \bar{d} i \gmf d 
- 4 m_s \bar{s} i \gmf s | p \ra
&=& g_{A}^{8} \cdot 2 M_{p} \, \bar{p} i \gmf p  \; , \\
 \la p | 2 m_u \bar{u} i \gmf u + 2 m_d \bar{d} i \gmf d 
+ 2 m_s \bar{s} i \gmf s | p \ra
&=& 0  \; . \nonumber 
\eea
We would like to stress that the last of Eqs.(\ref{14}) 
implies a resolution of the 
U(1) problem, i.e. the absence of a $ SU(3) $ flavor singlet 
meson with $ m^2 \sim m_q $. We are then able to find 
exact expressions for the $ SU(3) $
flavor dependent parts for each of the light flavor 
spin distributions separately:
\bea
\label{15}
\la p | 2 m_u \bar{u} i \gmf u | p \ra &=& \left( \frac{1}{2}
g_{A}^3 + \frac{1}{6} g_{A}^8 \right) 
\cdot 2 M_{p} \, \bar{p} i \gmf p 
= (0.725) \cdot 2 M_{p} \,  \bar{p} i \gmf p \; , \nonumber \\
\la p | 2 m_d \bar{d} i \gmf d | p \ra &=& \left( - \frac{1}{2}
g_{A}^3 + \frac{1}{6} g_{A}^8 \right) 
\cdot  2 M_{p} \, \bar{p} i \gmf p = 
( - 0.532) \cdot 2 M_{p}  \, \bar{p} i \gmf p \; , \\
\la p | 2 m_s \bar{s} i \gmf s | p \ra &=& \left( 
- \frac{1}{3} g_{A}^8 \right) \cdot 2 M_{p} \, 
\bar{p} i \gmf p = 
( - 0.193) \cdot 2 M_{p} \, \bar{p} i \gmf p \; . \nonumber
\eea
In the numerical values in (\ref{15}) we have implied that 
the numbers $ g_{A}^3 , g_{A}^8 $ in the chiral limit are not 
very different from their phenomenological values in the 
real world (\ref{5}). We expect an accuracy of this approximation
to be rather good\footnote{ More accurately, it can be shown that a 
decrease of $ M_{p} $ due to diminishing of the $s$-quark 
mass nearly compensates an increase of  $  g_{A}^3 , g_{A}^8 $
in the chiral limit. Thus, their product remains approximately
the same as in the real world.}. 
It is curious to note that, despite of their very transparent
meaning and exactness in
the chiral limit, we were unable to find relations (\ref{15}) in 
 the literature.
Meanwhile, they reveal something very interesting. Indeed, comparing 
these numbers to Ellis-Karliner fit of experimental results 
(\ref{8}) (remember that we have agreed 
to understand them as ``shifted" values $ \Delta q' = \Delta q + 
2/3 \Delta c $), we see that, taken separately, each light flavor
spin distribution is given by a large $ SU(3) $ dependent part 
plus a small $ SU(3) $ singlet piece whose
numerical value is approximately {\bf the same
and is very close to $ 0.1$  for all
the light flavors.} We thus see that the mystery of 
$ \Gamma_{1}^p
$ has to a large extent disappeared: the $ SU(3)$  flavor dependent 
parts give the 
main contributions to each of the light flavor spin distributions
$ \Delta q $, but, on the other hand, these $ SU(3) $ variant parts 
cancel out in the total light flavor 
contribution $ \Delta \Sigma $, as a consequence of resolution of 
the U(1) problem (see the last of Eqs.(\ref{14})). 

Therefore, if we managed to establish theoretically the number
\beq
\label{16}
\frac{1}{2 M_{p} \bar{p} i \gmf p }
\la p | \frac{ \alpha_s}{4 \pi} G \tilde{G}
- \frac{2}{3} \frac{1}{16 \pi^2 m_{c}^2 } g^3 G \tilde{G} G 
| p \ra \simeq 0.1 \; , 
\eeq
 (which is simply $1/3 \, g_{A}^0$, see Eq.(\ref{11})),
the $ \Gamma_{1}^{p}$ problem would be resolved in QCD terms.
Note that constraint (\ref{16}), resulting from comparison of 
numbers (\ref{15}) with Ellis-Karliner fit (\ref{8}), agrees
with Eq.(\ref{11}) obtained without a specification of different 
flavor contributions to the quantity $ \Sigma + 2 \Delta c $.
Thus, $ SU(3) $ singlet contribution (\ref{16}) is indeed flavor 
independent as should be expected. Furthermore, it 
consists of two parts. Had we knew the anomaly 
matrix element over the proton independently, we would 
then be able 
to extract the intrinsic charm component to the proton
spin (the second term in Eq.(\ref{16})) from  
experimental constraint (\ref{16}). As will be shown
in the next section, this knowledge is provided by the KZ theorem,
and thus such a separation can indeed be made. 

\section{K\"{u}hn-Zakharov low energy theorem and intrinsic
charm in the proton spin}

Our aim in this section is to separate the anomaly and intrinsic
charm contributions in Eq.(\ref{16}). To this end, we are going 
to a use a beautiful result by K\"{u}hn and Zakharov (KZ) \cite{KZ}
which provides us with an {\it exact} (in the chiral limit)
answer for the proton matrix element of the topological density
(the first term in (\ref{16})). KZ have shown that the famous 
dimensional transmutation phenomenon, which lies at heart of 
QCD low energy theorems, actually also fixes a value of the latter
matrix element. In view of the fact that this important result
is usually either ignored or doubted in the literature, we would 
like first to present KZ arguments at length and advocate their 
correctness. Next, we will use this result to obtain our first 
estimate of the intrinsic charm contribution to the proton spin.

We start with recalling  the well known fact that in a massless 
asymptotically free theory (such as QCD in the chiral limit)
the appearance of a mass scale is related to the so-called
dimensional transmutation phenomenon, according to which 
in this case the only 
mass parameter in the theory is 
\beq
\label{17}
m \equiv \Lambda \exp \left( - \frac{8 \pi^2}{ b g^2 ( \Lambda) } 
\right) \; ,   
\eeq 
where $ b = 11/3 N_c - 2/3 n_f $  
is the first coefficient of the Gell-Mann - Low 
beta function and $ \Lambda
$ stands for a ultraviolet cut-off ( $N_c $ is a number 
of colors). By construction, $ m $ does not depend on the cut-off
$ \Lambda $, and all physical parameters (masses, vacuum condensates,
etc.) in the chiral limit can only be proportional to a power
of the mass parameter (\ref{17}), as they cannot depend on the 
cut-off explicitly. Specifying on the proton mass $ M_{p} \sim m  $, 
this means that under the variation of the cut-off
\beq
\label{18}
\Lambda \rightarrow \Lambda ( 1 + \varepsilon)
\eeq
a total variation of $ M_{p} $ must vanish:
\beq
\label{19}
\varepsilon \Lambda \frac{d M_{p}}{ d \Lambda} = 0 \; .     
\eeq
On the other hand, the total derivative in respect 
to $ \Lambda $ can be represented
as the sum of a partial derivative and a term corresponding 
to a variation of the Lagrangian under cut-off shifts 
(\ref{18}) :
\beq
\label{20}
0 =  \varepsilon \Lambda \frac{d M_{p}}{ d \Lambda} =
\varepsilon \Lambda \frac{\partial M_{p}}{ \partial 
\Lambda} + \delta_{\Lambda} M_{p} \; , 
\eeq
where 
\beq
\label{21}
\delta_{\Lambda} M_{p} \, \bar{p} p = - \la p | \delta_{\Lambda}
L^{(\Lambda)} | p \ra = - \la p | \varepsilon \Lambda  
 \frac{\partial L^{(\Lambda)} }{ \partial 
\Lambda} | p \ra
\eeq
and $ L^{(\Lambda)} $ is a regularized QCD Lagrangian. Its variation
under scale transformations (\ref{18})  is expressed by the 
conformal anomaly equation through the trace of the energy-momentum
tensor $ \theta_{\mu \mu} $
\beq
\label{22}
 \varepsilon \Lambda  
  \frac{\partial L^{(\Lambda)} }{ \partial 
\Lambda} = \varepsilon \theta_{\mu \mu} = - \varepsilon
\frac{b \alpha_s}{ 8 \pi} G_{\mu \nu}^a G _{\mu \nu}^a  \; . 
\eeq 
On the other hand, the partial derivative of $ M_{p} $ is 
fixed by Eq.(\ref{17}) :
\beq
\label{23}
\varepsilon \Lambda \frac{\partial M_{p}}{ \partial 
\Lambda} = \varepsilon M_{p} \; .
\eeq
Combining Eqs.(\ref{20},\ref{21},\ref{22},\ref{23}), we obtain
\beq
\label{24}
\la p | - \frac{b \alpha_s }{ 8 \pi} G^2 | p \ra = M_{p} \, 
\bar{p} p \; . 
\eeq
Eq.(\ref{24}) is well known and expresses the fact that in the chiral
limit the proton mass is given by the conformal anomaly matrix 
element over the proton. As was pointed out in \cite{KZ}, 
Eq.(\ref{24}) can be obtained in a number of ways, thus providing
an independent check of the above procedure.

A particularly interesting observation made by KZ \cite{KZ} was 
that the very same line of reasoning could be applied in situations
when only a fermionic cut-off (specifically the mass $ M_{R} $
of a Pauli-Villars fermion regulator) is varied in two
possible ways:
\beq
\label{25}
M_{R} \rightarrow M_{R}( 1+ \varepsilon)
\eeq
or 
\beq
\label{26}
M_{R} \rightarrow M_{R} ( 1 + i \varepsilon) \; , 
\eeq 
where $ \varepsilon $ is a small and real number. 
Consider first variation 
(\ref{25}). Again, the renormalizability of the theory implies
that a total variation of the proton mass under shifts (\ref{25}) must 
vanish. On the other hand, a variation of the Lagrangian under
a real shift of $ M_R $  can be readily found. Indeed, at the one-loop
level contributions of fermion and boson regulators to the 
conformal anomaly are additive, and therefore the variation of the
Lagrangian under transformations (\ref{25}) is 
\beq
\label{27}
\delta_{M_{R}} L = \varepsilon \cdot \frac{2}{3} n_{f} 
\alpha_s G^2 \; , 
\eeq
which together with Eq.(\ref{24}) yields
\beq
\label{28}
\varepsilon M_{R} \frac{ \partial M_{p}}{
\partial M_{R} } = - \varepsilon \frac{2 n_f }{ 3 b } M_{p} \; .
\eeq
Here we come to a most interesting (and sometimes wrongly 
criticized
in the literature) part of the KZ proposal. Consider now 
the complex
regulator mass variation (\ref{26}). This variation leads to 
a $ \gmf $ mass term for the Pauli-Villars fermion $R$, 
as Eq.(\ref{26})
is equivalent to the transformation
\beq
\label{29}
M_{R} \bar{R} R \rightarrow 
M_{R} \bar{R} R +  \varepsilon 
M_{R} \bar{R} i \gmf R \; .
\eeq
Therefore, the variation of the Lagrangian in this case is 
proportional to the axial anomaly
\beq
\label{30}
\delta_{i M_{R}} L = \varepsilon n_f \frac{\alpha_s }{
8 \pi} G \tilde{G} \; .     
\eeq
Again, a shift in the physical proton mass due to the matrix 
element of anomaly (\ref{30}) must be cancelled by a partial 
derivative in respect to variation (\ref{26}) (see a 
comment below). On the other hand, the partial derivative     
is now determined by Eq.(\ref{28}) :
\beq
\label{31}
i \varepsilon  M_{R} \frac{ \partial M_{p}}{
\partial M_{R} } = - i \varepsilon 
\frac{2 n_f }{ 3 b } M_{p} \; .
\eeq
Therefore,  matrix element of the chiral anomaly
(\ref{30}) is fixed \cite{KZ}
by the requirement of independence 
of the physical mass of shifts (\ref{26}) :
\beq
\label{32}
\la p | \frac{ \alpha_s }{ 8 \pi} G \tilde{G} | p \ra 
= - \frac{2 }{ 3 b } M_{p} \, \bar{p} i \gmf p \; ,
\eeq
that completes the proof of the KZ theorem. As was stressed
in \cite{KZ}, the above derivation implies (again) a resolution of
the U(1) problem since otherwise the matrix element of 
interest would
not exist in the limit of vanishing momentum transfer because of
a massless U(1) goldstone contribution. Moreover, while there
may exist higher
order corrections to Eq.(\ref{32}), this result is exact in the 
chiral limit at least to the one-loop accuracy. Its validity was 
checked by KZ in the case of supersymmetric QCD, and may also be 
tested in solvable models. 

As the negative sign in Eq.(\ref{32}) is so important
for what follows, it
is instructive to have an alternative  derivation of this result,
where the sign would be explained in a  simple and intuitive way.
We now present such a derivation.

The physical meaning of Eq.(\ref{24}) is very simple: it 
says that in the chiral limit
a considerable part of the nucleon mass
$ \frac{11 N_c}{3b} \, M_p $ comes from gluons,
while the term related to quarks (more precisely, 
to their regulator fields)
is equal to $ \frac{-2n_f}{3b} \, M_p $.
Let us introduce chiral combinations of the regulator fields in 
the standard way:
\beq
\label{new3}
R_l=\frac{1}{2}(1+\gmf)R,~~R_r=\frac{1}{2}(1-\gmf)R .
\eeq
Transformation properties of the
$R_l,~R_r$ fields under chiral rotations 
$\sim \exp(i\alpha\gmf)$
 have the very simple form:
\beq
\label{new4}
\bar{R_r}R_l\rightarrow \exp(i\alpha)\bar{R_r}R_l,~~
\bar{R_l}R_r\rightarrow \exp(-i\alpha)\bar{R_l}R_r
\eeq
 
The requirement of reparametrization invariance 
under chiral rotations (\ref{new4}) 
   in the limit $m_q=0$ can be expressed by equations 
\beq
\label{new5}
\la p | -M_R\bar{R_r}R_l e^{i\alpha} | p \ra 
=   M_{p}^{(f)} e^{i\alpha}\, \bar{p}_r   p_l \; ,~~~~~~~~~~
\la p | -M_R\bar{R_l}R_r e^{-i\alpha} | p \ra 
=   M_{p}^{(f)} e^{-i\alpha}\, \bar{p}_l   p_r \; ,
\eeq
which should be valid for arbitrary $\alpha$. In this formula
  $M_{p}^{(f)}$ 
is a part 
 of the nucleon mass which comes from the fermion
regulator field\footnote{ Different $n_f$
fermions multiply this result by the factor $n_f$.}:
 $M_{p}^{(f)}=(\frac{-2}{3 b}) \, M_p
$. 
One can easily check this formula
 once again  using an expansion for the scalar density 
at  $M_{R} \rightarrow \infty$:
 \beq
\label{new6}
\la p |-M_R\bar{R }R| p \ra \rightarrow
\la p |  \frac{ \alpha_s }{ 12 \pi} G^2 +0(1/M_R^2)| p \ra 
=M_{p}^{(f)} \bar{p}p 
 = \frac{-2 }{3b} M_p \, \bar{p}p \; ,
\eeq
which is exactly the   contribution of one flavor
to complete expression (\ref{24}).  
A similar calculation for the pseudoscalar part of Eq.(\ref{new5}) 
gives the result (for convenience, we multiply both parts
of the equation by the factor $i$):
\beq
\label{new7}
\la p |-M_R\bar{R }i\gmf R| p \ra \rightarrow
 \la p |  \frac{ \alpha_s }{ 8 \pi} G \tilde{G}
   +0(1/M_R^2)| p \ra =M_{p}^{(f)}\bar{p}i\gmf p 
 = \frac{-2 }{3b} M_p \, \bar{p}i\gmf p,
\eeq
where we used the standard expansion 
$-M_R\bar{R }i\gmf R\rightarrow \frac{ \alpha_s }{ 8 \pi} 
G \tilde{G}$
for the large mass $M_R$, see Eq.(\ref{new1}).
Expression (\ref{new7}) is precisely KZ theorem (\ref{32}).
Now we understand the sign in KZ formulae (\ref{32}), (\ref{new7}) 
very well: it has the {\bf minus} sign because a fermion field
gives a negative contribution to the nucleon mass in 
comparison with
the main term originating from gluons. By the same reasons 
this term is suppressed
in the large $N_c$ limit.
This statement is very clear and unambiguous.

We would now like to guess    
a reason why the KZ formula is mostly ignored in the 
literature. 
The reason is 
the negative sign of matrix element (\ref{32}), which is 
absolutely crucial in what follows. Had we forgotten about 
the second
term in Eq.(\ref{16}), we would conclude that result (\ref{32}) 
contradicts the data as it is of a different sign. Our answer
 to this 
objection is that it is the second charm-induced term in 
Eq.(\ref{16})
that makes the whole expression positive. As will be shown 
in the 
next section, an independent estimate of this term,  
based on results on $ B \rightarrow \eta'$ decays, confirms 
this 
claim\footnote{ Here we would like to mention that in 
alternative approach
to the anomaly matrix element over the proton, which is 
based on U(1) 
Goldberger-Treiman relations within a ``two-component" 
approach of 
Ref.\cite{GT} (see also references in \cite{Ans}),  
a sign of matrix element (\ref{32}) comes as a 
result of interplay between the $ \eta' $ and a composite gluon 
ghost field interactions with the nucleon, which may in 
principle lead to  
any signature. Yet, in the model-independent KZ method this sign 
is fixed to 
be negative. A discussion of possible consequences of this 
observation
for the formalism of Ref.\cite{GT} would lead us far beyond the 
scope of this paper.}.      

Finally, we are ready to use the KZ theorem given 
by Eq.(\ref{32}) to find a 
charm-induced contribution to the proton spin from experimental 
constraint (\ref{16}). We obtain
\beq
\label{35}
\frac{1}{ 2 M_{p} \, \bar{p} i \gmf p} 
\la p | - \frac{1}{16 \pi^2 m_{c}^2 } \, g^3 G \tilde{G} G
| p \ra \simeq 0.3
\eeq
We thus conclude that experimental result (\ref{16}), taken together
with KZ formula (\ref{32}), requires a substantial contribution of 
intrinsic charm to the proton spin, which is about twice 
larger than 
the total {\bf singlet} light flavor contribution given by the 
first term in 
Eq.(\ref{16}). 
 We should note that, taken separately, the $u$ and $d$
quark contributions (\ref{15})  are numerically considerably 
larger than the $c$-quark term.
It is the singlet combination which is  
 smaller than the
  charmed quark contribution.
  Such a smallness of the singlet combination seems natural
  within the large $N_c$  approach: it should vanish
  according to KZ formula (\ref{32}). We consider  this 
  as a reasonable qualitative explanation of our result:  
     the charm contribution becomes competitive in comparison 
  with the $ SU(3) $ singlet light quark term because the latter 
is simply 
suppressed by the factor $1/N_c$ on a natural scale.  
    Nevertheless, we understand  that
this result may be not easy to reconcile with one's 
intuition. On the other hand, the first term in Eq.(\ref{16}) is 
the matrix element of a total derivative, i.e. a very unusual object.
The fact that it does not vanish can be
understood as resulting from interactions in a gauge-variant
sector of the theory. The physical intuition can hardly work
in such untypical situation. Thus, to support the above conclusion
on a role of intrinsic charm in the proton spin,
we need to find matrix element (\ref{35}) in an independent way.
If number (\ref{35}) was established without a reference to the 
polarized DIS data, the latter would be completely explained 
in terms of theoretically calculated matrix elements (\ref{32}) and 
(\ref{35}) (see Eq.(\ref{16}) above). 
We will now argue 
that estimate (\ref{35}) can be confirmed within a different line 
or reasoning, based on a very different experiment,
 which is completely independent of both constraint 
(\ref{16}) and KZ result (\ref{32}).

\section{ Intrinsic charm in the proton and $ B \rightarrow 
\eta' $ data}

We now proceed to an independent estimate of matrix element (\ref{35})
which measures a contribution of the intrinsic charm component of
the proton spin to the first moment $ \int dx g_{1}(x) $ of the
polarized structure function $ g_{1}(x) $. Somewhat unexpectedly, 
such an estimate can be obtained from  new data of 
B-physics. As will be argued in this section,  
experimental information
on $ B \rightarrow \eta' $ decays can be used to make conclusions 
on a role of the intrinsic charm component $ \la
p| \bar{c} \gmmu \gmf c | p \ra $ in the proton spin. 
We will show that this quantity may be related to the 
intrinsic charm
component $ \lo \bar{c} \gmmu \gmf c \er $ of the $ \eta'
 $. The latter
matrix element is actually  
probed experimentally as will be argued below. Thus, in what follows 
we will first spend some time discussing the B-physics, and then 
demonstrate how this can be helpful in the problem of interest.
   
Recently, CLEO collaboration has reported \cite{CLEO}
results of measurements of inclusive and exclusive production of 
the $ \eta' $ in B-decays :
\beq
\label{36}
Br( B \rightarrow \eta' + X \; ; 2.2 \; GeV < E_{\eta'} <
2.7 \; GeV ) = (7.5 \pm 1.5 \pm 1.1) \cdot 10^{-4} \; ,  
\eeq
\beq
\label{37}
Br(B \rightarrow  \eta'+ K ) = (7.8_{-2.2}^{+
2.7} \pm 1.0) \cdot 10^{-5} \; .
\eeq
Here the inclusive branching ratio contains the acceptance 
cut intended to reduce a background 
from events with charmed meson interactions in a final state. At 
first
sight, the above numbers might seem quite innocent. However, 
simple 
calculations \cite{1,2} reveal that these data are in severe 
contradiction 
with a standard view of the process at the 
quark level as a decay of the $b$-quark into 
 light quarks, which could be naively suggested  
 keeping in mind the standard picture of   
 $ \eta' $ as  a SU(3) singlet
meson made of the $u-$, $d-$ and $s-$quarks. In this picture 
the $ B \rightarrow \eta' $  amplitude must be proportional to the 
Cabbibo 
suppression
factor $V_{ub}$ and, as a result, the standard mechanism of 
the $ B \rightarrow \eta' $ transition yields numbers which are 
by two orders of magnitude (!) smaller than the data for both the 
inclusive and exclusive cases. Thus, there must
be something beyond this standard picture. Furthermore, 
inclusive decays
are usually dominated by few-particle final states. Therefore,
any meaningful mechanism, intended to explain abnormally large 
numbers (\ref{36},\ref{37}), should deal with both the inclusive 
and exclusive data at the same time.

A particular mechanism, suggesting a unified description of both
the inclusive and exclusive modes, was proposed in our recent 
papers \cite{1,2}. We have shown that at the quark level $ B 
\rightarrow \eta' $ decays can be described by the Cabbibo favored
$ b \rightarrow c \bar{c} s $ process followed by a transition
of $ \bar{c} c $ into the $ \eta' $. The latter transition 
is possible due to an intrinsic charm component of the $ \eta' $, 
and quantitatively can be expressed through the matrix element
\beq
\label{38}
\lo \bar{c} \gmmu \gmf c | \eta'(q) \ra \equiv i \fc q_{\mu} \; . 
\eeq
A reason why this matrix element may be non-zero can be explained 
as follows. As the $c$-quark is heavier than the $ \eta' $, it may
only exist in the $ \eta' $ in a loop. Such a loop 
with the heavy $c$-quark can be evaluated 
in terms of gluon fields, populating the $ \eta' $, by using the 
background field technique, see Eq.(\ref{new1}). Therefore, 
matrix element (\ref{38}) may not vanish due to $ \bar{c} c 
\leftrightarrow gluons $ transitions. 

Prior to a theoretical calculation of the charm current residue 
$ \fc $ into the $ \eta' $, it is instructive to find this quantity
``experimentally". By this we mean a value of $ \fc $ needed for the 
above mechanism to explain the data. Simple calculations show
that the number \cite{1,2}
\beq
\label{39} 
\fc \simeq 140 \; MeV \; \; (``exp")
\eeq 
yields a good agreement with the data in both the inclusive and
exclusive modes. On the other hand, this value may look 
uncomfortably large as it is only a few times smaller than the 
analogously normalized residue $ \lo \bar{c} \gmmu \gmf c 
| \eta_c (q) \ra = i f_{\eta_c} q_{\mu} $ with $ f_{\eta_c} 
\simeq 400 \; MeV $ known experimentally from the $ \eta_c 
\rightarrow \gamma \gamma $ decay. Intuitively, these numbers
could be expected to differ much more drastically as, in contrast
to $ f_{\eta_c} $, the residue $ \fc $ is a double 
suppressed amplitude. It is Zweig rule-violating
and besides contains the $1/m_c^2$ suppression factor, as 
it comes from loop effects.
 Therefore, one could expect it to be very small. 
In reality it is not small.
There are two reasons
for this. First, $m_c$ is not very large on the hadronic $1\; 
GeV$ scale. 
Second, and more important,
the Zweig rule itself is badly broken down in 
vacuum $0^{\pm}$ channels.  Of course, such a breakdown  
contradicts
a naive large $N_c$   
counting where a non-diagonal transitions
should be suppressed in comparison with 
 diagonal ones.
However, a more careful analysis \cite{NSVZ,1} reveals
that the large $ N_c $ picture and breakdown of the 
Zweig rule in fact peacefully co-exist: while the 
large $ N_c $ description is quite accurate for the 
$ \eta' $, an extent to which the Zweig rule is violated
in $ \eta' $ yields a large residue 
$ \fc $. We stress that 
the phenomenon of the breakdown of the Zweig rule 
in vacuum $ 0^{\pm} $ channels is well known and 
understood \cite{NSVZ}, and many phenomenological
examples of corresponding physics have been discussed in the 
literature, see e.g. \cite{NSVZ,arz}. 
The residue 
$\fc$  (which is fundamentally important 
for our estimates)
is another manifestation of the same physics.

We now proceed to a theoretical analysis of 
the residue $ \fc $ 
defined by Eq.(\ref{38}). Unfortunately, a detailed 
theoretical 
consideration of this 
quantity would require a repetition of our original paper \cite{1}
almost at full length. We thus refer to Ref.\cite{1} for details, 
while here we would like to give an idea and flavor of our method.
It is convenient to start with taking the derivative 
in Eq.(\ref{38}). Then we obtain
\beq
\label{40}
\fc  = \frac{1}{m_{\eta'}^2} \lo 2 m_c \bar{c} i \gmf c +
\atop \er \; . 
\eeq
 Since the $c$-quark is heavy, one can use the Operator Product
Expansion in inverse powers of the $c-$quark mass (the heavy 
quark expansion)   
\beq
\label{heavy}
2 m_c \bar{c} i \gmf c = - \atop - \frac{1}{16 \pi^2
 m_{c}^2 }
g^3 f^{abc} G_{\mu \nu}^a \tilde{G}_{\nu \alpha}^b 
G_{\alpha \mu}^c 
+ \ldots  
\eeq
(see the appendix in \cite{1} for a detailed derivation 
of this result). Further terms in expansion (\ref{heavy}) are 
neglected in what follows (see however below). 
We have thus reduced the problem to a calculation of the 
matrix element of the purely gluonic operator:
 \beq 
\label{42}
\fc = - \frac{1}{16 \pi^2 m_{\eta'}^2 } \frac{1}{m_{c}^2}
\lo g^3 f^{abc} G_{\mu \nu}^a \tilde{G}_{\nu \alpha}^b 
G_{\alpha \mu}^c \er \; . 
\eeq
It may seem at first sight that Eq.(\ref{42}) is of little
help since matrix elements of gluon operators
are difficult to calculate. A situation with 
the $ \eta'$ is, however, exceptional,
as the $ \eta' $ is strongly coupled to 
gluons and to some extent can be viewed as a remnant of 
imaginary purely gluonic world. In effect, matrix
element (\ref{42}) is amenable to a theoretical study
which essentially reduces to a nonperturbative analysis of 
pure Yang-Mills theory. 
Closely following old ideas due to 
Witten\cite{Witten} and Veneziano\cite{Ven} 
and using some additional (nonperturbative)
arguments, we have managed to estimate  matrix element
(\ref{42})  
\cite{1}:
\beq
\label{43}
\fc \simeq \frac{3}{4 \pi^2 b} \frac{1}{m_c^2} \frac{
\la g^3 G^3 \ra _{YM}}{ \lo \atop \er }  \; . 
\eeq
Therefore,   
  we have related the residue of the charmed 
axial current into the $ \eta'$ with apparently 
completely unrelated quantity which a  
cubic gluon condensate in the pure Yang-Mills theory
(we notice that the matrix element of topological density 
  which appears in
(\ref{43}) is known $ \lo (\alpha_s/ 4\pi) G \tilde{G}
\er \simeq 0.04 \; GeV^3 $ \cite{NSVZ}).
Using all currently available information 
regarding the vacuum condensate
$\la g^3 G^3 \ra _{YM}$ in gluodynamics,
we have arrived at the numerical estimate \cite{1}
\beq 
\label{44}
\fc = ( 50 - 180) \; MeV \; . 
\eeq
In spite of a large uncertainty of this result, its  main source 
is nevertheless well localized and 
related  to a poor knowledge of the 
cubic condensate in pure gluodynamics. One can see that our 
theoretical estimate (\ref{44}) agrees within errors with 
``experimental" value (\ref{39}).

By reasons mentioned above, it is extremely important to calculate
matrix element (\ref{42}) in an independent way within a 
QCD-based model in order to test the general idea.
Such a calculation based on the instanton-liquid model (see e.g. 
\cite{Shur} for a review 
and references to original papers)
has been carried out \cite{ShurZh} with the following
promising result: the matrix element
of the three-gluon operator as defined by Eq.(\ref{42})
is very close to ``experimental'' value (\ref{39}).
We should remind the reader that this model is extremely successful   
in description of low-energy hadronic properties, and a typical
accuracy of this approach is not worse than 30 \%.
A qualitative, microscopic explanation of the 
enhancement of matrix element (\ref{42}) is the following.
The model suggests the existence of relatively small 
strongly interacting instantons with a strong gluon field 
inside the instanton.
This field is so strong that  a relevant parameter
$gG/m_c^2 $ is not  small numerically
as one could naively expect, but rather is of order one. 
We are quite confident in calculations
\cite{ShurZh} because a similar calculation with the standard
gluon current $\tilde{G}G$ is in a good agreement with the existing
phenomenological data on the $\eta'$  \cite{Shur}.

We interpret all these results as the evidence that the suggested
mechanism indeed explains the data on $  B \rightarrow \eta' $ given 
by numbers (\ref{36},\ref{37}). Therefore, in what follows we 
consider the intrinsic charm component of the $ \eta' $, described
by Eqs.(\ref{38},\ref{39},\ref{43},\ref{44}), as an established 
property of the $ \eta' $ which is probed in experiments 
and, on the other hand, is understood theoretically.
Still, in view of a poor accuracy of our theoretical prediction
(\ref{44}), we will use ``experimental" number (\ref{39}) in 
our subsequent estimates.

We are now in a position to make our second estimate of an
intrinsic charm component of the proton spin. Consider the 
matrix element ($ q = p' - p $, other notations 
here are self-explanatory) 
\beq
\label{47}
\la N(p') | \bar{c} \gmmu \gmf c | N(p) \ra = 
\bar{u} (p') \left[  \gmmu \gmf h_{1}(q^2) + 
q_{\mu} \gmf h_{2} (q^2) \right] u(p) \; . 
\eeq
Taking the derivative and using Eq.(\ref{heavy}), we obtain
\beq
\label{48}
\la N(p') |  - \frac{1}{16 \pi^2 m_{c}^2 }
g^3 G \tilde{G} G  | N(p) \ra = i \bar{u} (p') \gmf u(p)
\left[  2 M_{N} h_{1}(q^2) + 
q^2  h_{2} (q^2) \right]  \; .
\eeq
Assuming now a $ \eta' $ dominance in this matrix 
element\footnote{Note that such a saturation becomes exact 
in the 
large $ N_c $ limit.} and 
using the absence of pole singularities in the
second form factor $ h_2 $, which ensures a resolution of the 
U(1) problem, we arrive at the Goldberger-Treiman type relation
\beq
\label{49}
h_{1}(0) = \frac{1}{2 M_{N}} g_{\eta' NN} \fc
\eeq 
or, equivalently,  
\beq
\label{50}
\frac{1} { 2 M_{N} \, \bar{u}(p) 
i \gmf u(p) } \, \la N(p) |  - \frac{1}{16 \pi^2 m_{c}^2 }
g^3 G \tilde{G} G  | N(p) \ra =   \frac{1}{2 M_{N}} 
g_{\eta' NN} \fc \; .
\eeq
Note that owing to Eq.(\ref{heavy}) the anomaly term cancels out in 
(\ref{50})\footnote{ This is in contrast to the standard case 
with U(1) 
Goldberger-Treiman relations for the $SU(3)$ singlet light flavor 
current discussed in Ref.\cite{GT}. The anomaly term is a total
derivative, and the fact that it gives a non-zero contribution to
the proton matrix element is due to the presence of 
non-pole subtraction
terms in dispersion relations, which have the same nature as 
subtraction terms in a correlation function of the topological   
density \cite{Ven}, and can be assigned to a ghost contribution.
This is the reason why the proton spin has a ``two-component" form 
in the approach of Ref.\cite{GT}. As a result, the light quark
 component of the proton spin is expressed in terms of two unknown
quantities, neither of which is directly measurable. For these 
reasons 
we prefer to use  the exact KZ theorem  for the anomaly 
matrix element rather than results from \cite{GT}. On the 
contrary to the case
\cite{GT},   
the anomaly term does not show up in 
Eq.(\ref{50}), and this is why it is the physical coupling constant
$ g_{\eta' NN } $ that stands there.}. Unfortunately, the 
precise value
of $ g_{\eta' NN } $ is not known, and phenomenological 
estimates of the coupling 
 $ g_{\eta' NN } $ vary from 
$ g_{\eta' NN } \simeq 3 
$ to $ g_{\eta' NN } \simeq 7 $ \cite{coup}. Assuming the
same range for this quantity in the chiral limit
and using numerical value 
(\ref{39}), we arrive at the estimate
\beq
\label{51}
 \frac{1} { 2 M_{N} \, \bar{u}(p) 
i \gmf u(p) } \, \la N(p) |  - \frac{1}{16 \pi^2 m_{c}^2 }
g^3 G \tilde{G} G  | N(p) \ra =   0.2 - 0.5  \; .
\eeq
This result has a large uncertainty mainly due to a poor 
knowledge of $ g_{\eta' NN} $. In view of a large uncertainty in
$ g_{\eta' NN} $, we have neglected a perturbative evolution of
$ \fc $.
 
A few remarks are in order. First, as  we already mentioned, 
the microscopic picture
suggests that the matrix element under consideration is not small 
because of gluon fluctuations which  could be large. In 
different words, 
the parameter of the heavy quark expansion $gG/m_c^2$ could 
be of order one. 
Such an explanation might imply a bad  
convergence
of expansion (\ref{heavy}) where we 
  keep the first term and neglect  all the rest. 
A scientific  answer to the question 
of convergence of series (\ref{heavy}) would require the 
knowledge of  
 higher dimensional matrix elements 
which are not available at the moment. Alternatively, one 
could try to address this issue
within the instanton liquid approach \cite{Shur}. However, 
irrespectively
to an outcome of such an analysis, we could consider 
``experimental" value of $ \fc $ (\ref{39}), which effectively
takes into account all powers of the $ 1/m_c $ expansion.  
Therefore,  all of them are also effectively included in the left
hand side of Eq.(\ref{51}).  Thus, a  possible
bad convergence of expansion (\ref{heavy}) does not affect
our results at all. 
 Hence 
we see that
our final estimate (\ref{51}) confirms the 
above conclusion on a large $c$-quark component in the proton 
spin, as its sign and order of magnitude agree with 
phenomenological constraint
(\ref{35}). The large magnitude of the charmed current 
residue $ \fc $ into the $ \eta' $ is the main factor 
giving rise to large number (\ref{51}). We emphasize again 
that estimate (\ref{51}) 
is obtained in an entirely independent of the DIS phenomenology
and KZ theorem way, and relies on the data and theoretical results 
of B-physics. If we fix the $ \eta' NN $ coupling 
constant in the chiral limit by ``experimental" constraint
(\ref{35}), it is consistent within uncertainties 
with independent theoretical 
prediction (\ref{51}). We thus conclude that the polarized 
DIS data agree within errors with theoretical results expressed 
by Eqs.(\ref{16}), (\ref{35}) and (\ref{51}). This implies 
that the main contribution to $ \Gamma_{1}^p $ is due to the 
charm component of the proton spin. We now proceed to a discussion
of possible ways to test this conclusion.   

\section{Signals for intrinsic charm in polarized experiments}
  
In this section we would like to argue that our
explanation of the polarized DIS data has very definite 
consequences for future polarized 
experiments which thus will be able to test the picture suggested by 
this paper. We will only deal with qualitative aspects of expected
phenomena, which follow immediately from our results. A more detailed
analysis is beyond the scope of this paper and deserves a separate
study.

We have found that the light quark and charm contributions
to the first moment $ \Gamma_{1}^p $ have different signs and 
magnitudes.   Here and in what follows any references to the 
proton spin are understood in the sense of 
the singlet part of the spin operator $ \Delta \Sigma  + 2 \Delta 
c $ only.
 While the former is negative, the latter is positive 
with about a twice larger magnitude. The immediate consequence 
of this 
fact is that an integral asymmetry for events with no charm in the 
final state is expected to be negative with approximately 
the same magnitude
as a total integral asymmetry. This prediction may hopefully be 
tested experimentally, provided the two types of final states can 
be distinguished.

Yet, a most interesting test of our scenario is related 
to 
a charm distribution in charmed final states. We find that the 
charm contribution to $ \Gamma_{1}^p $ is about an order of 
magnitude larger than results obtained within perturbative schemes
\cite{pert}. The difference from previous analyses comes due to 
strong nonperturbative fluctuations in the axial $SU(3) $
singlet channel, which lead to a large magnitude of 
the intrinsic charm component of the 
proton spin. Our definition of the 
latter quantity matches the well known definition \cite{intr} 
of the intrinsic 
charm in terms of a Fock state description as a multiple connected 
loop of the $c$-quark in the proton\footnote{See \cite{BK} for
a recent discussion of the intrinsic charm physics in a different
content.}. 
In our approach this property is used in a 
constructive way. Furthermore, one can expect that the intrinsic
charm component of the proton spin will be traced in polarized 
experiments. We note that in the case of unpolarized DIS various 
mechanisms, which are 
able to identify the intrinsic charm in the proton,
have been discussed in the literature. In particular, they include
a liberation of the charm as a result of  
scattering on light quarks \cite{BHMT,heavyDIS}, or 
direct DIS on the 
intrinsic charm \cite{Ing}. The intrinsic charm contribution to the 
proton spin, considered in this paper, corresponds 
to the latter case. 

There exists a clear distinction between pictures of a charm 
distribution in the final state corresponding to 
different mechanisms of the charm production in DIS. A vast 
literature \cite{PGF}
is devoted to the charm production in polarized experiments
as an effective tool to measure the gluon spin dependent 
distribution $ \Delta G(x, Q^2) $. This proposal is based on the 
perturbative photon-gluon fusion (PGF)
mechanism which assigns a production
of the $ \bar{c}c $ pair to the hard subprocess $ \gamma^{\ast} g 
\rightarrow \bar{c} c $. It is undoubtably true that the PGF 
process adequately describes a heavy quark production in the case of 
usual unpolarized DIS, and expresses it in terms of the unpolarized 
gluon distribution. Moreover, in the unpolarized case 
the manifestation of the  intrinsic charm component of the nucleon
is still an open problem \cite{Ing,heavyDIS}. We believe that 
the underlying reason for a relatively subdominant role 
\cite{HSV} of the 
intrinsic charm in usual DIS  
is a strong suppression of 
nonperturbative effects in vector channels. 
However, as we argued in Introduction, it would be potentially 
dangerous to automatically transfer concepts and results, valid 
for unpolarized DIS, to the polarized case. Although at the 
perturbative level results look similar, these arguments ignore
important differences between the two cases, which only show up 
beyond perturbation theory. That this way of thinking
is 
fallacious can be most clearly illustrated on the example of the 
pseudoscalar nonet. The Zweig rule is 100 \% violated there by 
nonperturbative effects, though no difference from the Zweig rule 
conserving case of vector mesons is seen in 
perturbation theory. 

As, in contrast to the perturbative PGF mechanism, polarized
DIS on the intrinsic charm does not have a strong $ \alpha_s $   
suppression, we expect that the charm production in 
polarized experiments will be overwhelmed by events due to  
intrinsic charm mechanisms. Quantitative predictions in this 
case are troublesome as, in addition to direct DIS on the intrinsic 
charm, there exist other possibilities for the extrinsic and intrinsic 
charm production in DIS \cite{BHMT,heavyDIS}. Model independent 
methods needed for their evaluation are not available at 
present, though
estimates with model light cone wave functions for the intrinsic 
charm \cite{intr} are possible. Nevertheless, different mechanisms 
for the charm production can be disentangled on the basis of a   
event topology analysis \cite{Ing,heavyDIS}.
 The PGF mechanism corresponds to two 
jets (plus a spectator) events 
with a charmed hadron (typically a $D$-meson) carrying  
a large fraction of the photon energy in the current jet and
an anticharm (e.g., $ \bar{D}$) hadron in the target jet. 
A topology of these PGF events is expected to deviate from the 
planar one \cite{Ing}. The $ \bar{c} c $ pair has a low invariant
mass as the fractional energy of the proton carried by the gluon
is typically small. The transverse momentum and invariant mass of 
the pair will grow with the photon virtuality $ Q^2 $. 
As has been argued above, we expect that these events will be 
parametrically suppressed by powers of $ \alpha_s $ in comparison
to events reviving the intrinsic charm in the nucleon.

As for DIS on the intrinsic charm, one can distinguish between two 
types of processes. The first one is an indirect process when 
the photon scatters on a light quark with a subsequent liberation of
the charm \cite{BHMT,heavyDIS}. 
In this case both the charm quark and antiquark are in the 
target fragmentation region, and can hadronize into both open charm
(e.g. $ D , \Lambda_c $) or hidden 
charm ($ J/ \psi $, etc.) hadrons.
In this process, the intrinsic charm shows up through a 
nonperturbative final state interaction. Charmonium states, 
produced in this region, will presumably have a large fraction
of the proton momentum with $ p_{\perp} \sim m_c $ and independent 
of $ Q^2 $. On general grounds, one may expect an excessive 
production of S-wave quarkonia (e.g. $ \eta_c $) in comparison 
to P-wave states.
 
The second process is direct DIS
on the intrinsic charm quark \cite{Ing}. The target charm 
quark is emitted 
at large forward rapidities, while the scattered quark has
somewhat smaller near forward rapidities. The difference in 
rapidities grows with $ Q^2 $. The average transverse momentum
of the $ \bar{c} c $ pair is not expected to grow with $ Q^2 $. 
Thus, our results allow one to suggest that a substantial 
(presumably dominant) part of charm events in 
polarized experiments at large $ Q^2 $ will be an open charm
hadrons (or S-wave quarkonia) production with low $ p_{\perp}
\sim m_c $.

Finally, we note that all said above on the charm production
can be extended to beauty production. The latter is 
suppressed relatively to the former by the factor $ m_{b}^2 / 
m_{c}^2 \simeq 10 $, which may imply that the beauty can be 
produced nonperturbatively via the intrinsic beauty mechanism 
at approximately the same level as the charm production through
the perturbative PGF process.  

\section{Conclusions}

In this paper we have presented a somewhat unorthodox point 
of view of the data on the first moment $ \int dx g_{1}(x) $
of the spin $ g_{1}(x) $ structure function
measured in polarized DIS experiments. We have abandoned 
attempts
to explain the data within the quark model  or related to it field
theoretical models, and reformulated the problem in terms of 
particular matrix elements of quark and gluon operators.  
In our opinion, none of the phenomena, which 
are of crucial importance for understanding the proton
spin problem (spontaneous breaking of the chiral invariance, a 
resolution
of the U(1) problem,  the dimensional transmutation, etc.), can 
be adequately described within the parton or quark model. 
On the contrary, 
all of them are essentially nonperturbative, and need to be 
addressed within nonperturbative QCD. We have further shown 
that the data can be interpreted as a constraint on the sum of 
matrix elements of the anomaly and intrinsic charm operators.    
The latter object is a new principal element of our analysis,
which has not been considered hitherto in the literature in 
the content of the proton spin problem. Still, we have argued that  
the intrinsic charm component of the 
proton spin is quite large and constitutes a main contribution   
to the first moment $ \Gamma_{1}^p $. Its large magnitude is 
related to strong nonperturbative fluctuations in the axial
$ SU(3) $ singlet channel. Moreover, the inclusion of the 
intrinsic charm proton spin in
the analysis is absolutely necessary as it is the 
only way to reconcile the polarized DIS data with both an exact 
K\"{u}hn-Zakharov low energy theorem on one side, and new 
data and theoretical results on $ B \rightarrow \eta' $ decays on 
the other. We believe that these seemingly uncorrelated 
problems are actually tightly connected.
The situation reminds us the $ J /\psi $ discovery in 1974, when
a charmonium state (``hidden charm") was observed simultaneously
in $ e^{+} e^{-} $ collisions at SLAC \cite{Augustin} and at the 
proton machine at Brookhaven \cite{Aubert}. We believe that we 
are now facing a similar case, when different experimental groups
see the ``intrinsic charm" in polarized DIS (\ref{10}) and 
B-decays (\ref{36},\ref{37}) simultaneously. 

 We feel that  
at a microscopic level, the new effect
 of large intrinsic charm fluctuations
 originates in the instanton physics \cite{Shur},  where strong
 gluon fields in the singlet channel are able to  lead to  
  very unexpected phenomena. Our results seem to imply a kind 
of universality of this nonperturbative physics with universality
classes defined by quantum numbers relevant for a process considered,
but not concrete particles involved.
  
 Our explanation of the 
polarized DIS data has very definite consequences for the future 
polarized experiments, and thus can be tested there.    
  
  \section*{Acknowledgments}  

We are very grateful to E. Shuryak 
 for numerous
 stimulating discussions and sharing with us his insight into 
instantons.
 A.R.Z. thanks
  Isaac Newton  Institute for Mathematical Sciences, Cambridge, UK  
   for its hospitality and partial
 support
during his visit in May-June 1997 which initiated the collaborative work
\cite{ShurZh}.

\end{document}